\documentclass[aps,prl,twocolumn,showpacs,groupedaddress]{revtex4}  % for review and submission
\usepackage{graphicx}  % needed for figures
\usepackage{dcolumn}   % needed for some tables
\usepackage{bm}        % for math
\usepackage{amssymb}   % for math
\usepackage{rotating}
\newcommand{\MET}   {\mbox{$\not \!\! E_T$}}

\newcommand{\ttbar} {t\bar{t}}

\newcommand{\bbar}  {\bar{b}}

\newcommand{\shat}  {$\sqrt{\hat{s}}$}
\newcommand{\mshat}  {\sqrt{\hat{s}}}

\def        \dzero  {D\O~}
\newcommand{\lsim}{\mathrel{\hbox{\rlap{\lower.55ex\hbox{$\sim$}} \kern-.3em 
\raise.4ex \hbox{$<$}}}}

\begin{document}
%

% the following line is for submission 
\hspace{5.2in} \mbox{Fermilab-Pub-06/257-E}

\title{Search for $W'$~boson production in the $W' \rightarrow t\bbar$ decay channel}

% LIST_OF_AUTHORS_R2.TEX                 6/19/06            
%
\author{                                                                      
%% names begin here                                                           
V.M.~Abazov,$^{36}$                                                           
B.~Abbott,$^{76}$                                                             
M.~Abolins,$^{66}$                                                            
B.S.~Acharya,$^{29}$                                                          
M.~Adams,$^{52}$                                                              
T.~Adams,$^{50}$                                                              
M.~Agelou,$^{18}$                                                             
J.-L.~Agram,$^{19}$                                                           
S.H.~Ahn,$^{31}$                                                              
M.~Ahsan,$^{60}$
% ...
\\                                                                            
\vskip 0.30cm                                                                 
\centerline{(D\O\ Collaboration)}                                             
\vskip 0.30cm                                                                 
}                                                                             
\affiliation{                                                                 
\centerline{$^{1}$Universidad de Buenos Aires, Buenos Aires, Argentina}       
\centerline{$^{2}$LAFEX, Centro Brasileiro de Pesquisas F{\'\i}sicas,         
                  Rio de Janeiro, Brazil}                                     
\centerline{$^{3}$Universidade do Estado do Rio de Janeiro,                   
                  Rio de Janeiro, Brazil}                                     
\centerline{$^{4}$Instituto de F\'{\i}sica Te\'orica, Universidade            
                  Estadual Paulista, S\~ao Paulo, Brazil}                     
\centerline{$^{5}$University of Alberta, Edmonton, Alberta, Canada,           
                  Simon Fraser University, Burnaby, British Columbia, Canada,}
\centerline{York University, Toronto, Ontario, Canada, and                    
                  McGill University, Montreal, Quebec, Canada}                
\centerline{$^{6}$Institute of High Energy Physics, Beijing,                  
                  People's Republic of China}                                 
\centerline{$^{7}$University of Science and Technology of China, Hefei,       
                  People's Republic of China}                                 
\centerline{$^{8}$Universidad de los Andes, Bogot\'{a}, Colombia}             
\centerline{$^{9}$Center for Particle Physics, Charles University,            
                  Prague, Czech Republic}                                     
\centerline{$^{10}$Czech Technical University, Prague, Czech Republic}        
\centerline{$^{11}$Center for Particle Physics, Institute of Physics,         
                   Academy of Sciences of the Czech Republic,                 
                   Prague, Czech Republic}                                    
\centerline{$^{12}$Universidad San Francisco de Quito, Quito, Ecuador}        
\centerline{$^{13}$Laboratoire de Physique Corpusculaire, IN2P3-CNRS,         
                   Universit\'e Blaise Pascal, Clermont-Ferrand, France}      
\centerline{$^{14}$Laboratoire de Physique Subatomique et de Cosmologie,      
                   IN2P3-CNRS, Universite de Grenoble 1, Grenoble, France}    
\centerline{$^{15}$CPPM, IN2P3-CNRS, Universit\'e de la M\'editerran\'ee,     
                   Marseille, France}                                         
\centerline{$^{16}$IN2P3-CNRS, Laboratoire de l'Acc\'el\'erateur              
                   Lin\'eaire, Orsay, France}                                 
\centerline{$^{17}$LPNHE, IN2P3-CNRS, Universit\'es Paris VI and VII,         
                   Paris, France}                                             
\centerline{$^{18}$DAPNIA/Service de Physique des Particules, CEA, Saclay,    
                   France}                                                    
\centerline{$^{19}$IPHC, IN2P3-CNRS, Universit\'e Louis Pasteur, Strasbourg,  
                    France, and Universit\'e de Haute Alsace,                 
                    Mulhouse, France}                                         
\centerline{$^{20}$Institut de Physique Nucl\'eaire de Lyon, IN2P3-CNRS,      
                   Universit\'e Claude Bernard, Villeurbanne, France}         
\centerline{$^{21}$III. Physikalisches Institut A, RWTH Aachen,               
                   Aachen, Germany}                                           
\centerline{$^{22}$Physikalisches Institut, Universit{\"a}t Bonn,             
                   Bonn, Germany}                                             
\centerline{$^{23}$Physikalisches Institut, Universit{\"a}t Freiburg,         
                   Freiburg, Germany}                                         
\centerline{$^{24}$Institut f{\"u}r Physik, Universit{\"a}t Mainz,            
                   Mainz, Germany}                                            
\centerline{$^{25}$Ludwig-Maximilians-Universit{\"a}t M{\"u}nchen,            
                   M{\"u}nchen, Germany}                                      
\centerline{$^{26}$Fachbereich Physik, University of Wuppertal,               
                   Wuppertal, Germany}                                        
\centerline{$^{27}$Panjab University, Chandigarh, India}                      
\centerline{$^{28}$Delhi University, Delhi, India}                            
\centerline{$^{29}$Tata Institute of Fundamental Research, Mumbai, India}     
\centerline{$^{30}$University College Dublin, Dublin, Ireland}                
\centerline{$^{31}$Korea Detector Laboratory, Korea University,               
                   Seoul, Korea}                                              
\centerline{$^{32}$SungKyunKwan University, Suwon, Korea}                     
\centerline{$^{33}$CINVESTAV, Mexico City, Mexico}                            
\centerline{$^{34}$FOM-Institute NIKHEF and University of                     
                   Amsterdam/NIKHEF, Amsterdam, The Netherlands}              
\centerline{$^{35}$Radboud University Nijmegen/NIKHEF, Nijmegen, The          
                  Netherlands}                                                
\centerline{$^{36}$Joint Institute for Nuclear Research, Dubna, Russia}       
\centerline{$^{37}$Institute for Theoretical and Experimental Physics,        
                   Moscow, Russia}                                            
\centerline{$^{38}$Moscow State University, Moscow, Russia}                   
\centerline{$^{39}$Institute for High Energy Physics, Protvino, Russia}       
\centerline{$^{40}$Petersburg Nuclear Physics Institute,                      
                   St. Petersburg, Russia}                                    
\centerline{$^{41}$Lund University, Lund, Sweden, Royal Institute of          
                   Technology and Stockholm University, Stockholm,            
                   Sweden, and}                                               
\centerline{Uppsala University, Uppsala, Sweden}                              
\centerline{$^{42}$Physik Institut der Universit{\"a}t Z{\"u}rich,            
                   Z{\"u}rich, Switzerland}                                   
\centerline{$^{43}$Lancaster University, Lancaster, United Kingdom}           
\centerline{$^{44}$Imperial College, London, United Kingdom}                  
\centerline{$^{45}$University of Manchester, Manchester, United Kingdom}      
\centerline{$^{46}$University of Arizona, Tucson, Arizona 85721, USA}         
\centerline{$^{47}$Lawrence Berkeley National Laboratory and University of    
                   California, Berkeley, California 94720, USA}               
\centerline{$^{48}$California State University, Fresno, California 93740, USA}
\centerline{$^{49}$University of California, Riverside, California 92521, USA}
\centerline{$^{50}$Florida State University, Tallahassee, Florida 32306, USA} 
\centerline{$^{51}$Fermi National Accelerator Laboratory,                     
            Batavia, Illinois 60510, USA}                                     
\centerline{$^{52}$University of Illinois at Chicago,                         
            Chicago, Illinois 60607, USA}                                     
\centerline{$^{53}$Northern Illinois University, DeKalb, Illinois 60115, USA} 
\centerline{$^{54}$Northwestern University, Evanston, Illinois 60208, USA}    
\centerline{$^{55}$Indiana University, Bloomington, Indiana 47405, USA}       
\centerline{$^{56}$University of Notre Dame, Notre Dame, Indiana 46556, USA}  
\centerline{$^{57}$Purdue University Calumet, Hammond, Indiana 46323, USA}    
\centerline{$^{58}$Iowa State University, Ames, Iowa 50011, USA}              
\centerline{$^{59}$University of Kansas, Lawrence, Kansas 66045, USA}         
\centerline{$^{60}$Kansas State University, Manhattan, Kansas 66506, USA}     
\centerline{$^{61}$Louisiana Tech University, Ruston, Louisiana 71272, USA}   
\centerline{$^{62}$University of Maryland, College Park, Maryland 20742, USA} 
\centerline{$^{63}$Boston University, Boston, Massachusetts 02215, USA}       
\centerline{$^{64}$Northeastern University, Boston, Massachusetts 02115, USA} 
\centerline{$^{65}$University of Michigan, Ann Arbor, Michigan 48109, USA}    
\centerline{$^{66}$Michigan State University,                                 
            East Lansing, Michigan 48824, USA}                                
\centerline{$^{67}$University of Mississippi,                                 
            University, Mississippi 38677, USA}                               
\centerline{$^{68}$University of Nebraska, Lincoln, Nebraska 68588, USA}      
\centerline{$^{69}$Princeton University, Princeton, New Jersey 08544, USA}    
\centerline{$^{70}$State University of New York, Buffalo, New York 14260, USA}
\centerline{$^{71}$Columbia University, New York, New York 10027, USA}        
\centerline{$^{72}$University of Rochester, Rochester, New York 14627, USA}   
\centerline{$^{73}$State University of New York,                              
            Stony Brook, New York 11794, USA}                                 
\centerline{$^{74}$Brookhaven National Laboratory, Upton, New York 11973, USA}
\centerline{$^{75}$Langston University, Langston, Oklahoma 73050, USA}        
\centerline{$^{76}$University of Oklahoma, Norman, Oklahoma 73019, USA}       
\centerline{$^{77}$Oklahoma State University, Stillwater, Oklahoma 74078, USA}
\centerline{$^{78}$Brown University, Providence, Rhode Island 02912, USA}     
\centerline{$^{79}$University of Texas, Arlington, Texas 76019, USA}          
\centerline{$^{80}$Southern Methodist University, Dallas, Texas 75275, USA}   
\centerline{$^{81}$Rice University, Houston, Texas 77005, USA}                
\centerline{$^{82}$University of Virginia, Charlottesville,                   
            Virginia 22901, USA}                                              
\centerline{$^{83}$University of Washington, Seattle, Washington 98195, USA}  
}                                                                             
%end                                                                          
  % input Dzero author list

\date{August 31, 2006}

\begin{abstract}
We present a search for the production of a new heavy gauge boson $W'$ that
decays to a top quark and a bottom quark. We have analyzed 230 pb$^{-1}$
of data collected with the \dzero detector at the Fermilab Tevatron
collider at a center-of-mass energy of 1.96~TeV. 
No significant excess of events above the standard model expectation
is found in any region of the final
state invariant mass distribution. We set upper limits on the production
cross section of $W'$~bosons times branching ratio to top quarks
at the 95\% confidence level for several different $W'$~boson masses. 
We exclude masses between 200~GeV and 610~GeV for a 
$W'$~boson with standard-model-like couplings, between 200~GeV and 
630~GeV for a $W'$~boson with right-handed couplings that is 
allowed to decay to both leptons and quarks, and between 200~GeV
and 670~GeV for a $W'$~boson with
right-handed couplings that is only allowed to decay to quarks.
\end{abstract}

\maketitle

\clearpage
%%%%%%%%%%%%%%%%%%%%%%%%%%%%%%%%%%%%%%%%%%%%%%%%%%%%%%%%%%%%
%%%%%%%BEGINNING OF DOCUMENT
%%%%%%%%%%%%%%%%%%%%%%%%%%%%%%%%%%%%%%%%%%%%%%%%%%%%%%%%%%%%
\normalsize

\vfill

%
%%%%%%%%%%%%%%%%%%%%%%%%%%%%%%%%%%%%%%%%%%%%%%%%%%%%%%%%%%%%%%%%%%%%%%%%%%
% Introduction
%%%%%%%%%%%%%%%%%%%%%%%%%%%%%%%%%%%%%%%%%%%%%%%%%%%%%%%%%%%%%%%%%%%%%%%%%%
%\section{Introduction}
%\label{sec:intro}
%\vspace{-0.4cm}

% Paragraph 1
The top quark sector offers great potential to look for new physics related to electroweak
symmetry breaking. In particular, it is a sensitive probe for the
presence of additional gauge bosons beyond those of the standard model (SM). 
Such new gauge bosons typically arise in extensions
to the SM from the presence of additional symmetry groups~\cite{Tait:2000sh,Sullivan:2002jt}.

% Paragraph 2
Direct searches for the production of additional heavy gauge bosons
have focused on the lepton final state of the $W'$~ boson decay
which has good  separation between the $W'$~boson signal
and the SM backgrounds. The $W'$~boson lower mass limit in this
decay channel is 786~GeV~\cite{Affolder:2001gr}. In these studies, the
$W'$~boson is allowed to have right-handed interactions with leptons
and quarks, and it is assumed that the right-handed neutrino is
lighter than the $W'$~boson. It is also possible that
such a $W'$~boson does not interact with leptons and neutrinos but only
with quarks. Searching in the quark decay channel avoids assumptions about the
mass of a possible right-handed neutrino.
Previous direct searches for $W'$~bosons in the quark decay channel
have excluded the mass range below 261~GeV~\cite{Alitti:1993pn}
and between 300~GeV and 420~GeV~\cite{Abe:1997hm}. Assuming that the 
$W'$~boson decays only to quarks and not to leptons yields a lower mass
limit of 800~GeV~\cite{Abazov:2003tj}. 
A search has also been performed in the single top quark final state of the
$W'$-boson decay. 
Assuming the $W'$~boson has only right-handed
interactions and does not decay to leptons, the lower limit on the
$W'$~boson mass is 566~GeV~\cite{Acosta:2002nu}. The comprehensive search presented here
includes all of these $W'$~boson models.
Indirect searches for evidence of a $W'$~boson depend on exactly 
how it interferes with the SM $W$~boson and the results are thus highly 
model specific (see Ref.~\cite{Sullivan:2002jt} and references therein).

% Paragraph 3
The top quark was discovered in 1995 by the CDF and \dzero collaborations~\cite{topdiscovery},
but the production of single top quark has not yet been observed.
Both collaborations have searched for single top 
quark production~\cite{Acosta:2001un,RunII:cdf_result,d0runI,Abazov:2005zz,singletop_prd}. 
At the $95\%$ confidence level, the upper limit measured by \dzero on the $s$-channel 
process is 6.4~pb, and
the limit measured by CDF is 13.6~pb. At the same confidence level, the limit on the
$t$-channel production cross section is 5.0~pb from \dzero and 10.1~pb
from CDF. For comparison, the next-to-leading order (NLO) SM single top quark 
production cross sections are
0.88~pb in the $s$-channel and 1.98~pb in the $t$-channel~\cite{sintop-xsecs}.

% Paragraph 4
The single top quark final state is especially sensitive to the presence of
an additional heavy boson, owing to the decay chain $W' \rightarrow t \bbar$, where the top
quark decays to a $b$~quark and a SM $W$~boson. This decay is kinematically 
allowed as long as the
$W'$ mass is larger than the sum of top and bottom quark masses, i.e. as long
as it is above about 200~GeV.

\begin{figure}[!h!tbp]
\begin{center}
\includegraphics[width=0.48\textwidth]
{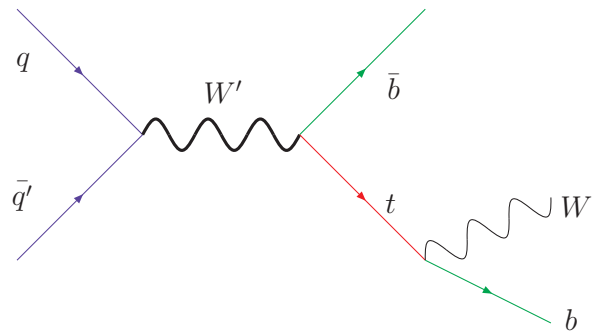}
\end{center}
\vspace{-0.5cm}
\caption{Leading order Feynman diagram for single top quark production
via a heavy $W'$~boson. The top quark decays to a SM $W$~boson and a $b$~quark.}
\label{fig:wprimediagram}
\end{figure}

% Paragraph 5
An additional heavy boson would appear as a peak in the invariant mass distribution
of the $t\bbar$ final state. Note that in this letter, the notation $t\bbar$ includes 
both final states $W'^{\;+} \rightarrow t\bbar$ and $W'^{\;-} \rightarrow \bar{t}b$.
The leading order Feynman diagram for $W'$~boson
production resulting in single top quark events is shown in Fig.~\ref{fig:wprimediagram}.
This diagram is identical to that for SM $s$-channel single top quark production
where the SM $W$~boson appears as the virtual 
particle~\cite{sintop-xsecs,Sullivan:2004ie,Campbell:2004ch,Cao:2004ap}.

% paragraph 6
The $W'$~boson also has a $t$-channel exchange that
leads to a single top quark final state. However, the cross section for a
$t$-channel $W'$ process is much smaller than the SM $t$-channel single top quark production
due to the high mass of the $W'$~boson. It will thus not be considered in this letter. 

% Paragraph 7
The SM $W$~boson from the top quark decay then
decays leptonically or hadronically. A heavy $W'$~boson could also contribute to the 
top quark decay, but that contribution is negligible, again because of the large 
$W'$~boson mass, and will not be considered here.

% Paragraph 8
We investigate three models of $W'$~boson production. In each case, we
set the CKM mixing matrix elements for the $W'$~boson equal to the SM values. 
In the first model ($W'_L$), we make the assumption that the coupling of the 
$W'$~boson to SM fermions is identical to that of the SM $W$~boson. 
Under these assumptions, 
there is interference between the SM $s$-channel single top quark process and 
the $W'$~boson production process from Fig.~\ref{fig:wprimediagram}. 
This interference term is small for large $W'$~boson masses, but it becomes
important in the invariant mass range of a few hundred GeV where the SM $s$-channel
production cross section is largest. 
In our modeling
of the $W'$~boson production process, we take this interference into
account. This is the first direct search for $W'$~boson production to
do so.

% Paragraph 9
In the second and third model ($W'_R$), the $W'$~boson has only right-handed
interactions, hence there is no interference with the SM $W$~boson.
In the second model, the $W'_R$~boson is allowed
to decay both to leptons and quarks, whereas in the third model
it is only allowed to decay to quarks. The main difference
between these two models is in the production cross section
and the branching fraction to quarks, and we use
the same simulated event sample for both models.
\begin{figure}[!h!tbp]
\begin{center}
\includegraphics[width=0.48\textwidth]
{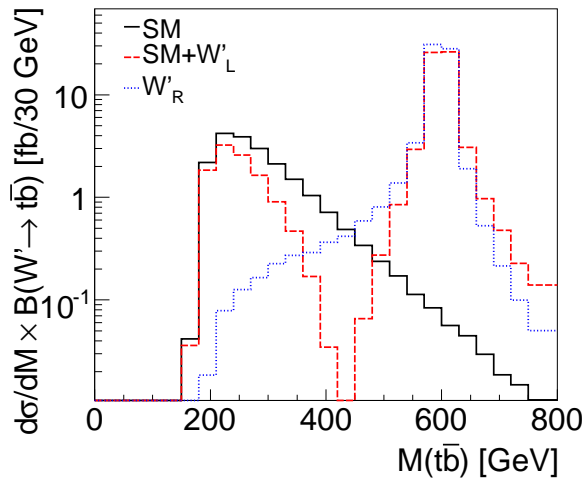}
\end{center}
\vspace{-0.5cm}
\caption{Histogram of the invariant mass of the top-bottom quark system
at the parton level for different models of $W'$~boson 
production. Shown are the SM $s$-channel distribution, the $W'_L\rightarrow t \bbar$~boson 
distribution, 
including the interference with the SM contribution, and the 
$W'_R\rightarrow t \bbar$~boson contribution, 
for a $W'$~boson mass of 600~GeV.}
\label{fig:shatcompare}
\end{figure}

% Paragraph 10
Figure~\ref{fig:shatcompare} compares the invariant mass distribution
for the $W'$~models with left-handed coupling (including interference)
and right-handed coupling (no interference) with the SM $s$-channel single top quark
distribution. While the position and width of the resonance peak at 600~GeV
is not very much affected by the interference, 
there is significant destructive interference for the left-handed coupling
in the invariant mass region between the SM and the resonance peak.

% Paragraph 11
Table~\ref{tab:xstable} shows the NLO 
cross sections for single top quark production
through a $W'$~boson for the three different models. The cross section
for SM-like left-handed $W'$~boson interactions takes into account the 
$W'_L$~boson contribution, the SM $s$-channel single top quark
contribution, and the interference between them. This combined cross section has been
calculated at leading order using {\sc CompHEP}~\cite{comphepref} and then multiplied by
the NLO/LO cross section ratio from Table~VII of Ref.~\cite{Sullivan:2002jt}. 
The factorization scale has been set equal to the invariant mass of the 
$W'$~boson.
There is no such interference term for right-handed
$W'$~boson interactions, and the cross sections in the two right
columns of Table~\ref{tab:xstable} have been taken directly from
Ref.~\cite{Sullivan:2002jt}. For $W'_R$~boson interactions,
the product of production cross section and branching fraction depends on whether the 
decay to leptons is allowed or not. The branching fraction for the decay
$W' \rightarrow t\bar{b}$ is about $3/12$ ($3/9$) if the $W'$~boson decay to quarks and leptons 
(only the decay to quarks) is allowed. 
The systematic uncertainty on the cross section includes components for
factorization and renormalization scale, top quark mass, and parton distribution
functions, and varies between about 12\% at a mass of
600~GeV and 18\% at a mass of 800~GeV.

\begin{table}[!h!tbp]
\begin{center}
\caption{Production cross section at NLO for a $W'$~boson $\times$ branching
fraction to $t\bar{b}$, for three different $W'$~boson models. The 
production cross sections for $W'_L$~boson
interactions also include the SM $s$-channel contribution as well as the interference
term between the two. They have been computed at leading order and scaled to NLO
according to Ref.~\cite{Sullivan:2002jt}. The cross sections for
$W'_R$~boson interactions differ depending on
which decays of the $W'$~boson are allowed. 
}
\label{tab:xstable}
\begin{ruledtabular}
\begin{tabular}{cccc}
$W'$ mass & \multicolumn{3}{c}{Cross section $\times$ $B$($W'\rightarrow t\bbar$)
[pb]} \\ 
 ~[GeV] &  SM+$W'_L$             & $W'_R$ ($\rightarrow$
$l$ or $q$) & $W'_R$ ($\rightarrow q$  only) \\
\hline
600         &  2.17   &  2.10    &  2.79 \\ 
650         &  1.43   &  1.25    &  1.65 \\ 
700         &  1.03   &  0.74    &  0.97 \\  
750         &  0.76   &  0.44    &  0.57 \\  
800         &  0.65   &  0.26    &  0.34 \\  
\end{tabular}
\end{ruledtabular}
\end{center}
\end{table}

% Paragraph 12
This analysis focuses on the final state topology of single top
quark production where the top quark decays into a $b$~quark and a 
SM $W$~boson, which subsequently decays leptonically ($W\rightarrow e\nu,~\mu
\nu$; including $W\rightarrow \tau\nu$ with $\tau \rightarrow e\nu,~\mu\nu$). This gives rise to an event signature with a high transverse
momentum lepton and significant missing transverse energy from the
neutrino, in association with two $b$-quark jets. The largest
backgrounds to this event signature come from $W$+jets and $\ttbar$
production. We also consider SM $t$-channel single top quark production
as a background in this search.

% Paragraph 13
The theoretical $W'$~boson production cross section is more than 15~pb for 
masses between 200~GeV and 
400~GeV for all three models considered here~\cite{Sullivan:2002jt}. 
The current limits on the single top quark
production cross section in the $s$-channel are
6.4~pb~\cite{Abazov:2005zz,singletop_prd} and 13.6~pb~\cite{RunII:cdf_result}
and don't depend much on whether the $W$~boson coupling is 
left-handed or right-handed.
Thus, $W'$~boson production with a decay to a top and a bottom quark
is excluded in this mass region. 
In this analysis we therefore explore the region of even higher masses.

% Paragraph 14
The analysis utilizes the same dataset, basic 
event selection, and background modeling as the \dzero single top quark
search described in Ref.~\cite{Abazov:2005zz}.
We select signal-like events and
separate the data into independent analysis sets based on final-state
lepton flavor (electron or muon) and $b$-tag multiplicity (single tagged or
double tagged), where $b$-quark jets are tagged using reconstructed
displaced vertices in the jets. The independent datasets are later
combined in the final statistical analysis. 
We perform a binned likelihood analysis on the invariant mass distribution
of all final state objects to obtain upper cross section limits at discrete
$W'$ mass points. We then compare these limits to
the theoretical prediction and derive a lower limit on the mass of the
$W'$~boson for each of the models under consideration.

% Paragraph 15
%%%%%%%%%%%%%%%%%%%%%%%%%%%%%%%%%%%%%%%%%%%%%%%%%%%%%%%%%%%%%%%%%%%%%%%%%%
% Detector
%%%%%%%%%%%%%%%%%%%%%%%%%%%%%%%%%%%%%%%%%%%%%%%%%%%%%%%%%%%%%%%%%%%%%%%%%%
%\section{The \dzero Detector}
%\vspace{-0.4cm}
The data for this analysis were recorded with the \dzero detector at the Fermilab
Tevatron, a 1.96~TeV proton-antiproton collider.
The \dzero detector has a central-tracking system, consisting of a 
silicon microstrip tracker and a central fiber tracker, 
both located within a 2~T superconducting solenoidal 
magnet~\cite{D0detector}, with designs optimized for tracking and 
vertexing at pseudorapidities $|\eta|<3$ and
$|\eta|<2.5$~\cite{fiducial_endnote}, respectively.
A liquid-argon and uranium calorimeter has a 
central section covering pseudorapidities $|\eta| \lsim 1.1$, and two end calorimeters that extend coverage 
to $|\eta|\approx 4.2$, with all three housed in separate 
cryostats~\cite{run1det}. An outer muon system, at $|\eta|<2$, 
consists of a layer of tracking detectors and scintillation trigger 
counters in front of 1.8~T iron toroids, followed by two similar layers 
after the toroids~\cite{run2muon}.

% Paragraph 16
%%%%%%%%%%%%%%%%%%%%%%%%%%%%%%%%%%%%%%%%%%%%%%%%%%%%%%%%%%%%%%%%%%%%%%%%%%
% Event Selection
%%%%%%%%%%%%%%%%%%%%%%%%%%%%%%%%%%%%%%%%%%%%%%%%%%%%%%%%%%%%%%%%%%%%%%%%%%
%\section{Data and Event Selection}
%\vspace{-0.4cm}
The analysis uses data recorded
between August 2002 and March 2004 ($230 \pm 15$~pb$^{-1}$ of integrated
luminosity). The data were collected using a
trigger that required an electromagnetic energy cluster and a jet in the
calorimeter for the electron channel, and
a muon and a jet for the muon channel. The event selection follows
that in Ref.~\cite{Abazov:2005zz}, except that only events with
two or three jets are allowed; four-jet events are excluded to reduce
the background contribution from $\ttbar$ production.

% Paragraph 17
In the electron channel, candidate events are selected by requiring
exactly one isolated electron (based on a seven-variable likelihood)
with transverse energy $E_{T} > 15$~GeV and $|\eta_{\rm det}|< 1.1$. In the muon
channel, events are selected by requiring exactly one isolated muon
with transverse momentum $p_{T} > 15$~GeV and $|\eta_{\rm det}|<2.0$. For both channels,
the events are also required to have missing transverse energy
$\MET>15$~GeV. Jets are required to have $E_{T}>15$~GeV and $|\eta_{\rm det}|<3.4$.
Events must have exactly two or exactly three jets, with the
leading jet additionally required to have $E_{T}>25$~GeV and
$|\eta_{\rm det}|<2.5$. 
At least one of the jets is required to be $b$-tagged using a secondary-vertex
algorithm~\cite{Abazov:2005ey}. We separate the dataset into
orthogonal subsets based on whether one or two jets are $b$-tagged.

% Paragraph 18
%%%%%%%%%%%%%%%%%%%%%%%%%%%%%%%%%%%%%%%%%%%%%%%%%%%%%%%%%%%%%%%%%%%%%%%%%%
% Selection Results, Background Modeling
%%%%%%%%%%%%%%%%%%%%%%%%%%%%%%%%%%%%%%%%%%%%%%%%%%%%%%%%%%%%%%%%%%%%%%%%%%
%\section{Signal and Background Modeling}
%\vspace{-0.4cm}
We estimate the acceptances for $W'$~boson production of single top quarks
using events generated by the \mbox{ {\sc CompHEP} 4.4.3} matrix element
event generator~\cite{comphepref}. The same program is also used to
estimate the yield for the SM single top quark background. Interference
between the SM $s$-channel and $W'_L$~boson production is taken into account
in the \mbox{\sc CompHEP} event generation for left-handed couplings.
The $W'$~boson signals
are normalized to the NLO cross section from Table~\ref{tab:xstable},
and we use the CTEQ6L1 parton distribution functions~\cite{Pumplin:2002vw}.

% Paragraph 19
We use both Monte Carlo and data to estimate the other background
yields. The $W$+jets and diboson ($WW$ and $WZ$) backgrounds are
estimated using simulated events generated with {\sc
alpgen}~\cite{Mangano:2002ea}. The diboson background yield is
normalized to NLO cross sections computed with {\sc MCFM}~\cite{mcfmref}. 
The fraction of
heavy-flavor ($Wb\bar{b}$) events in the $W$+jets background
is determined at the parton level, using {\sc MCFM} with the same 
parton-level cuts applied as for the samples used in the simulation. 
The overall $W$+jets yield is normalized to the
data sample before requiring a $b$-tagged jet. This normalization to data also
accounts for smaller contributions such as $Z$+jets events, where one
of the leptons from the $Z$~boson decay is not reconstructed. The
{$\ttbar$} background is estimated using simulated samples generated
with {\sc alpgen}, normalized to the (N)NLO cross section calculation:
$\sigma(\ttbar)=6.7 \pm 1.2$~pb~\cite{Kidonakis:2003qe}. The background
due to SM $t$-channel single top quark production is normalized to the 
NLO cross section calculation: 
$\sigma(tqb)=1.98 \pm 0.32$~pb~\cite{sintop-xsecs}. When investigating the
right-handed $W'$~boson coupling, the SM $s$-channel is also added as a background.
The uncertainty on the top quark mass is taken into account in the cross section
uncertainty.
The parton-level samples are then processed with {\sc
pythia 6.2}~\cite{pythiaref}
and a {\sc geant}~\cite{geantref}-based simulation of the \dzero
detector, and the resulting lepton and jet energies are further smeared to
reproduce the resolutions observed in data. Both the shape and the
overall normalization of the multijet background
is estimated from data, using multijet data samples that pass all
event selection cuts but fail the electron likelihood requirement in
the electron channel or the muon isolation requirement in the muon
channel.
The simulated signal and background samples include not only the decays
$W \rightarrow e\nu, \mu\nu$, but also the small contribution from
$W\rightarrow \tau\nu$ with $\tau \rightarrow e\nu,~\mu\nu$.

% Paragraph 20
%%%%%%%%%%%%%%%%%%%%%%%%%%%%%%%%%%%%%%%%%%%%%%%%%%%%%%%%%%%%%%%%%%%%%%%%%%
% Final Analysis
%%%%%%%%%%%%%%%%%%%%%%%%%%%%%%%%%%%%%%%%%%%%%%%%%%%%%%%%%%%%%%%%%%%%%%%%%%
%\section{Final Analysis}
%\vspace{-0.4cm}
The large mass of the $W'$~boson sets it apart from all background processes,
hence the best place to look for such a particle is the distribution of the 
reconstructed invariant mass in the resonance production process. 
We reconstruct the invariant mass of the $W'$~boson (the invariant mass of all final
state objects \shat) by adding the four-vectors 
of all reconstructed final state objects: the jets, the
lepton, and the neutrino from the $W$~boson decay from the top quark decay.
The $xy$-components of the neutrino momentum are given by the missing
transverse energy. The $z$-component is
calculated using a SM $W$~boson mass constraint, choosing the solution
with smaller $|p^\nu_z|$ from the two possible solutions. 
In order to isolate the $W'$~boson signal, we require $\mshat>400$~GeV.

\begin{figure}[!h!tbp]
\begin{center}
\includegraphics[width=0.48\textwidth]
{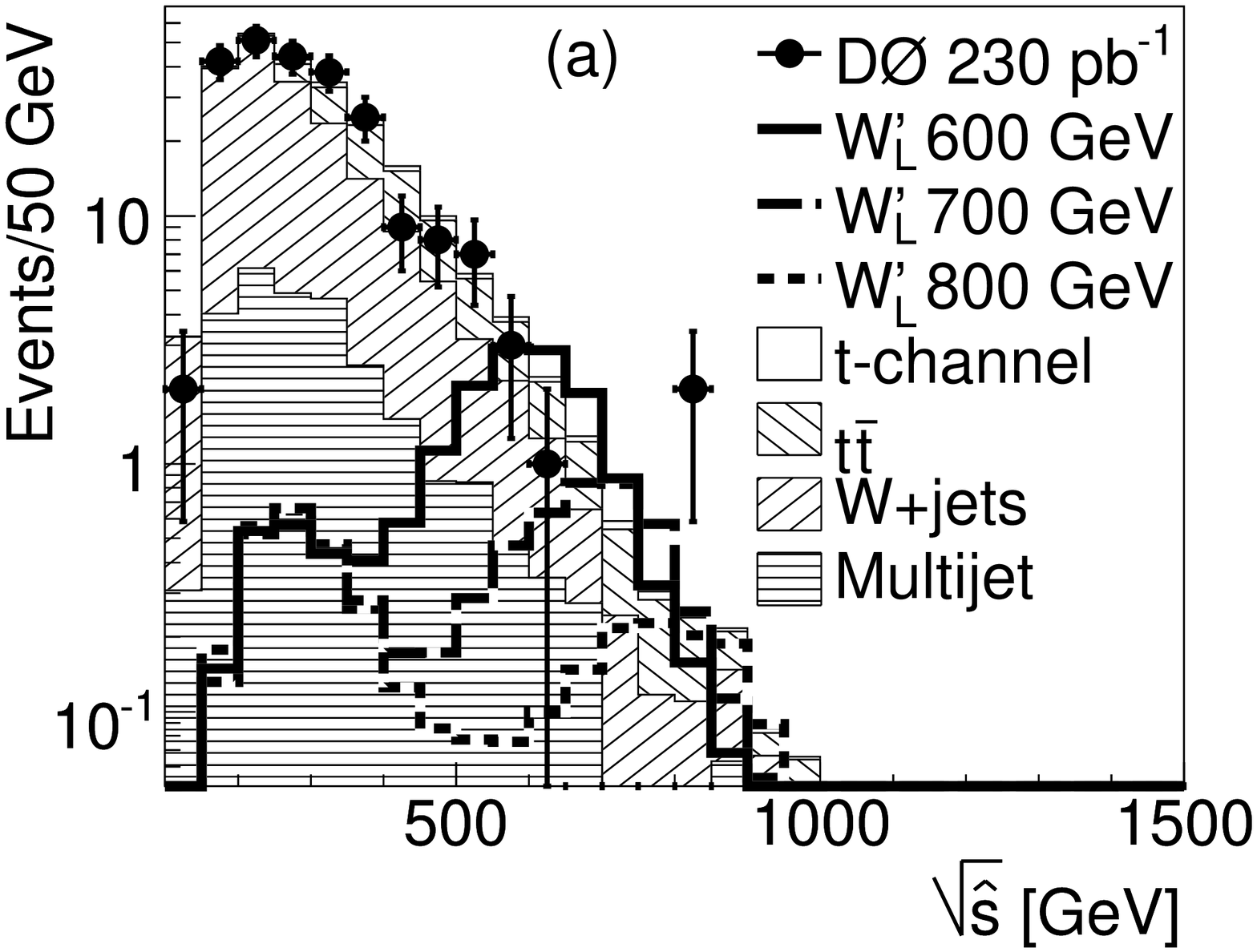}
\includegraphics[width=0.48\textwidth]
{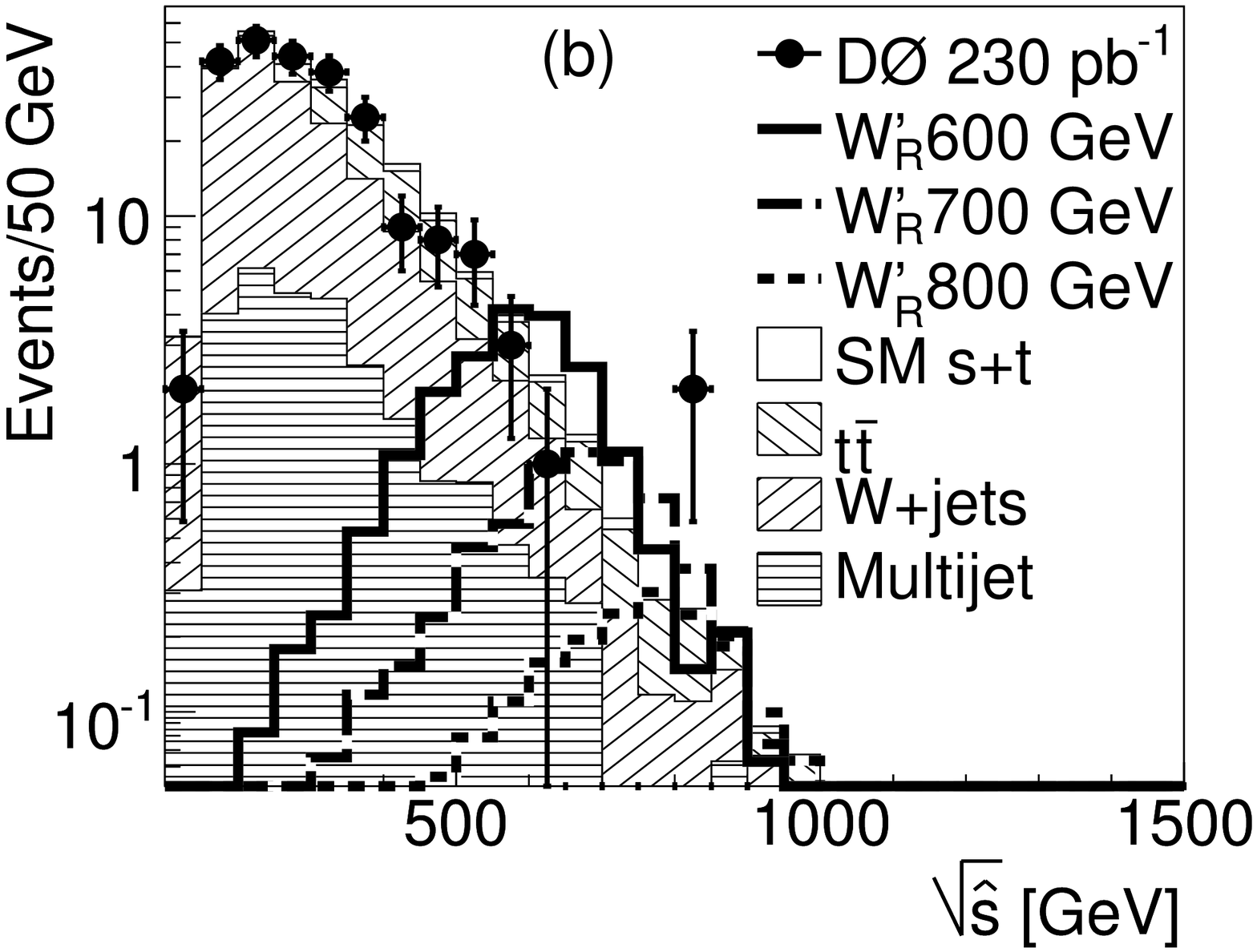}
\end{center}
\vspace{-0.5cm}
\caption{The reconstructed $W'$~boson invariant mass for several
different $W'$~boson masses as well as background processes for (a)
left-handed $W'$~boson couplings, and (b) right-handed couplings when only
the decay to quarks is allowed. Electron, muon,
single-tagged, and double-tagged events are combined. }
\label{fig:shatdata}
\end{figure}

% Paragraph 21
Figure~\ref{fig:shatdata} shows a comparison of the invariant mass distribution in data
to the sum of all background processes. Also shown are the expected contributions for
$W'$~bosons with left-handed and right-handed couplings at three different masses.

% Paragraph 22
The observed event yield is consistent with the background model in
every bin within uncertainties. There are two events at an invariant
mass of more than 800~GeV, with an expected
background of about 0.5 events. This excess of events is
consistent with an upward fluctuation of the background.

% Paragraph 23
%%%%%%%%%%%%%%%%%%%%%%%%%%%%%%%%%%%%%%%%%%%%%%%%%%%%%%%%%%%%%%%%%%%%%%%%%%
% Final Analysis
%%%%%%%%%%%%%%%%%%%%%%%%%%%%%%%%%%%%%%%%%%%%%%%%%%%%%%%%%%%%%%%%%%%%%%%%%%
%\section{Systematic Uncertainties}
%\vspace{-0.4cm}
Systematic uncertainties are evaluated for the simulated signal
and background samples, separately for electrons and muons and for each 
$b$-tag multiplicity. 
The dominant sources of systematic uncertainty on the signal and
background acceptances are (a) the uncertainty on the $b$-tag modeling
in the simulation, (b) the uncertainty from the jet energy scale, (c)
5\% uncertainty on the object identification efficiencies, (d)
5\% uncertainty on the trigger modeling, and (e) 5\%
uncertainty on the modeling of jet fragmentation~\cite{singletop_prd}. 
Each of these systematic uncertainties has been evaluated by varying the uncertainty 
for each object in the event (electrons, muons, jets) up and down by 
one standard deviation, and then propagating the updated objects and 
corresponding weights through the analysis chain.
The uncertainty on
the integrated luminosity is 6.5\%. 
%~\cite{Edwards:2004jz}. 
The background yields also have uncertainties from the cross sections, 
which vary from 8\% for diboson production to 15\% for SM
$t$-channel single top quark production and 18\% for the 
$\ttbar$ samples~\cite{Kidonakis:2003qe}. 
Since the $W$+jets background is normalized to the data before tagging, the
yield estimate is mainly affected by uncertainties related to
$b$-tagging. These include the $b$-tag modeling uncertainty, and the
uncertainty in the flavor composition before tagging derived from {\sc MCFM}, which is
estimated at 25\%. The $W$+jets background yield estimate also has an uncertainty
component from the parton level modeling of the \shat~distribution, 
which we estimate as 10\% based on 
event yield comparisons in the sample before requiring a $b$-tag.
The uncertainty in the background yield due to the jet energy scale 
varies between 15\% and 30\% for the single top, top pair, and diboson background samples. 
The uncertainty is large in these samples because most events have a small
invariant mass and only very few events
are in the region $\mshat>400$~GeV. Changing the jet energy by a small amount doesn't change
the overall distribution very much, but it has a large
impact on the number of events in the region $\mshat>400$~GeV.
The uncertainty from $b$-tag modeling is
about 8\% in the single-tagged sample and about 20\% in the double-tagged one.
The total uncertainty on the multijet samples is large ($\approx35\%$) due to 
the small number of 
events in the data sample used to model this background.

% Paragraph 24
Due to their similar kinematic properties, the $W'$~boson signal 
processes all have very similar systematic uncertainties.
The overall yield uncertainty due to the jet energy scale is small (1--2\%) 
for the signal processes because most of the signal events are in the region 
$\mshat>400$~GeV. 
The overall yield uncertainty for the signal samples has significant contributions 
from $b$-tag modeling (4\% for the single-tagged, 16\% for the double-tagged sample) 
and trigger modeling. 
The uncertainty in the signal region is significantly larger.
For example, the yield uncertainty due the jet energy scale for a cut of $\mshat>600$~GeV 
is about 40\% for the $W'_R$~(600~GeV) sample.

% Paragraph 25
Table~\ref{tab:yield_shat} shows the event yield
in the region $\mshat>400$~GeV
for all samples, including the total systematic uncertainty.
The uncertainty includes both acceptance and normalization components.

\begin{table}[!h!tbp]
\begin{center}
\caption{Event yields with uncertainty after selection, for the electron and muon
channel, single-tagged and double-tagged samples combined,
after event selection and requiring $\mshat>400$~GeV. The $W$+jets row also
includes diboson backgrounds. The total uncertainty on the background sum
takes correlations between different backgrounds into account.}
\label{tab:yield_shat}
\begin{ruledtabular}
\begin{tabular}{lccc}
 & \multicolumn{3}{c}{\underline{Event Yields for $\mshat>400$~GeV} }\\
      &  SM+$W'_L$             & $W'_R$ ($\rightarrow$
$l$ or $q$) & $W'_R$ ($\rightarrow q$  only) \\
\hline
Signals                  &        \\
~$W'$ (600~GeV)    & 13.0 $\pm 2.3$  & 13.8 $\pm 2.4$   & 18.4 $\pm 3.2$  \\
~$W'$ (650~GeV)    &  7.1 $\pm 1.3$  &  7.9 $\pm 1.1$   & 10.4 $\pm 1.5$  \\
~$W'$ (700~GeV)    &  4.4 $\pm 0.8$  &  4.6 $\pm 0.8$   &  6.0 $\pm 1.1$   \\
~$W'$ (750~GeV)    &  2.4 $\pm 0.4$  &  2.6 $\pm 0.5$   &  3.4 $\pm 0.6$   \\
~$W'$ (800~GeV)    &  1.6 $\pm 0.3$  &  1.5 $\pm 0.3$   &  1.9 $\pm 0.4$   
\vspace{0.02 in}                                          \\
Backgrounds         &         \\
~SM $t$-channel    &  \multicolumn{3}{c}{ 1.9 $\pm 0.8$  }\\
~${\ttbar}$        & \multicolumn{3}{c}{16.9  $\pm 5.6$  }\\
~$W$+jets          & \multicolumn{3}{c}{17.8  $\pm 4.5$  }\\
~Multijet          & \multicolumn{3}{c}{4.4   $\pm 1.5$  }
\vspace{0.02 in}                                          \\
Background sum      & \multicolumn{3}{c}{41.0 $\pm 10.2$ } \\
\hline
Data                     & \multicolumn{3}{c}{30 }
\end{tabular}
\end{ruledtabular}
\end{center}
\end{table}

% Paragraph 26
%%%%%%%%%%%%%%%%%%%%%%%%%%%%%%%%%%%%%%%%%%%%%%%%%%%%%%%%%%%%%%%%%%%%%%%%%%
% Limit setting procedure
%%%%%%%%%%%%%%%%%%%%%%%%%%%%%%%%%%%%%%%%%%%%%%%%%%%%%%%%%%%%%%%%%%%%%%%%%%
%\section{Cross Section Limits}
%\vspace{-0.4cm}
The observed data are consistent with the background predictions within
uncertainties. We therefore set upper limits on the $W'$~boson
production cross section for several different $W'$~boson masses in
each model. We use a Bayesian approach~\cite{IainTM2000} and
follow the formalism given in Ref.~\cite{Abazov:2005zz}.
The limits are derived from a likelihood function that 
is proportional to the probability to obtain the number of observed 
counts. Binned likelihoods are formed
based on the final state invariant mass distribution, assuming
a Poisson distribution for the observed counts and
a flat prior probability for the signal cross section. The priors for
the signal acceptance and the background yields are multivariate
Gaussians centered on their estimates and described by a covariance
matrix taking into account correlations across the
different sources and bins.

\begin{figure}[!h!tbp]
\begin{center}
\includegraphics[width=0.48\textwidth]
{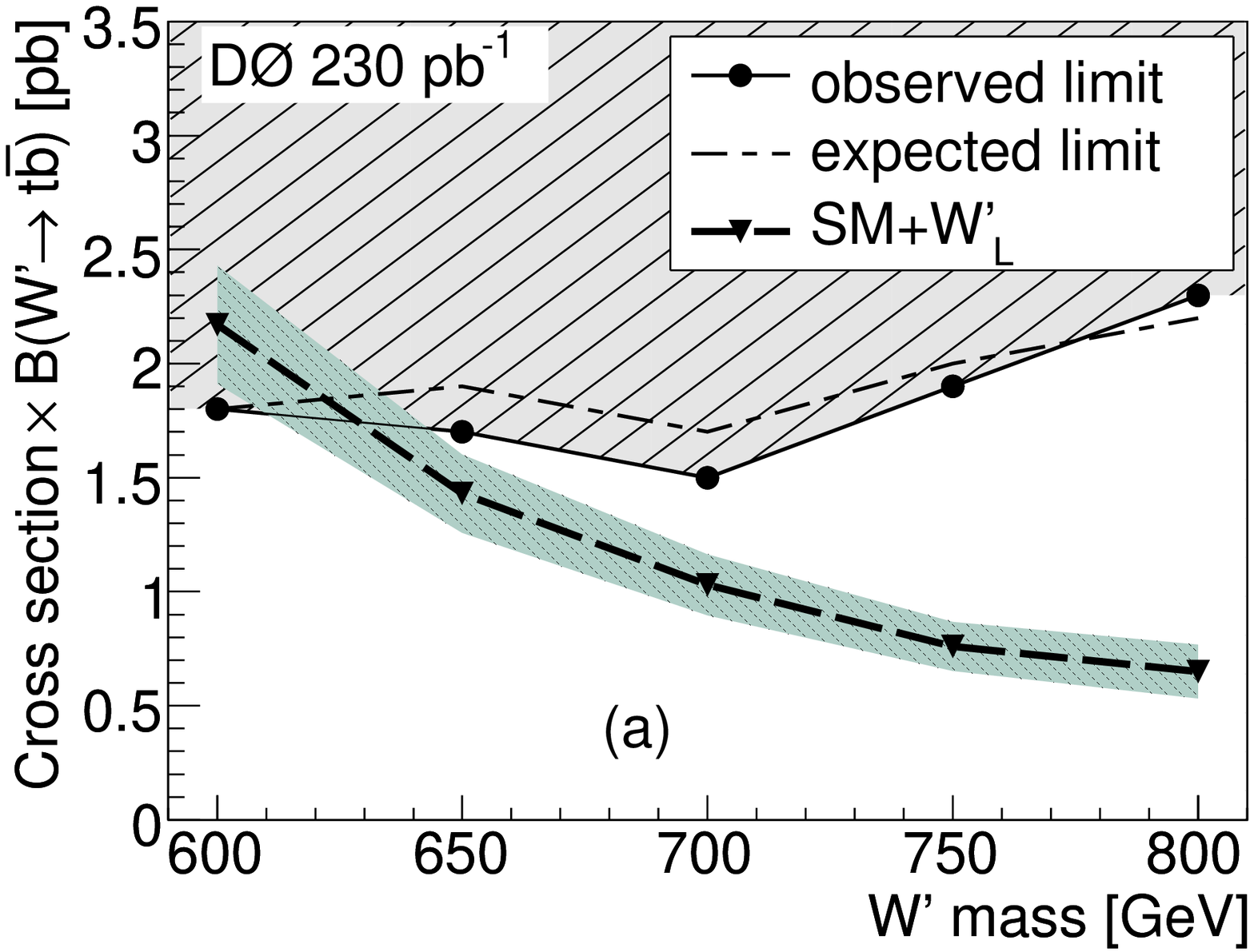}
\includegraphics[width=0.48\textwidth]
{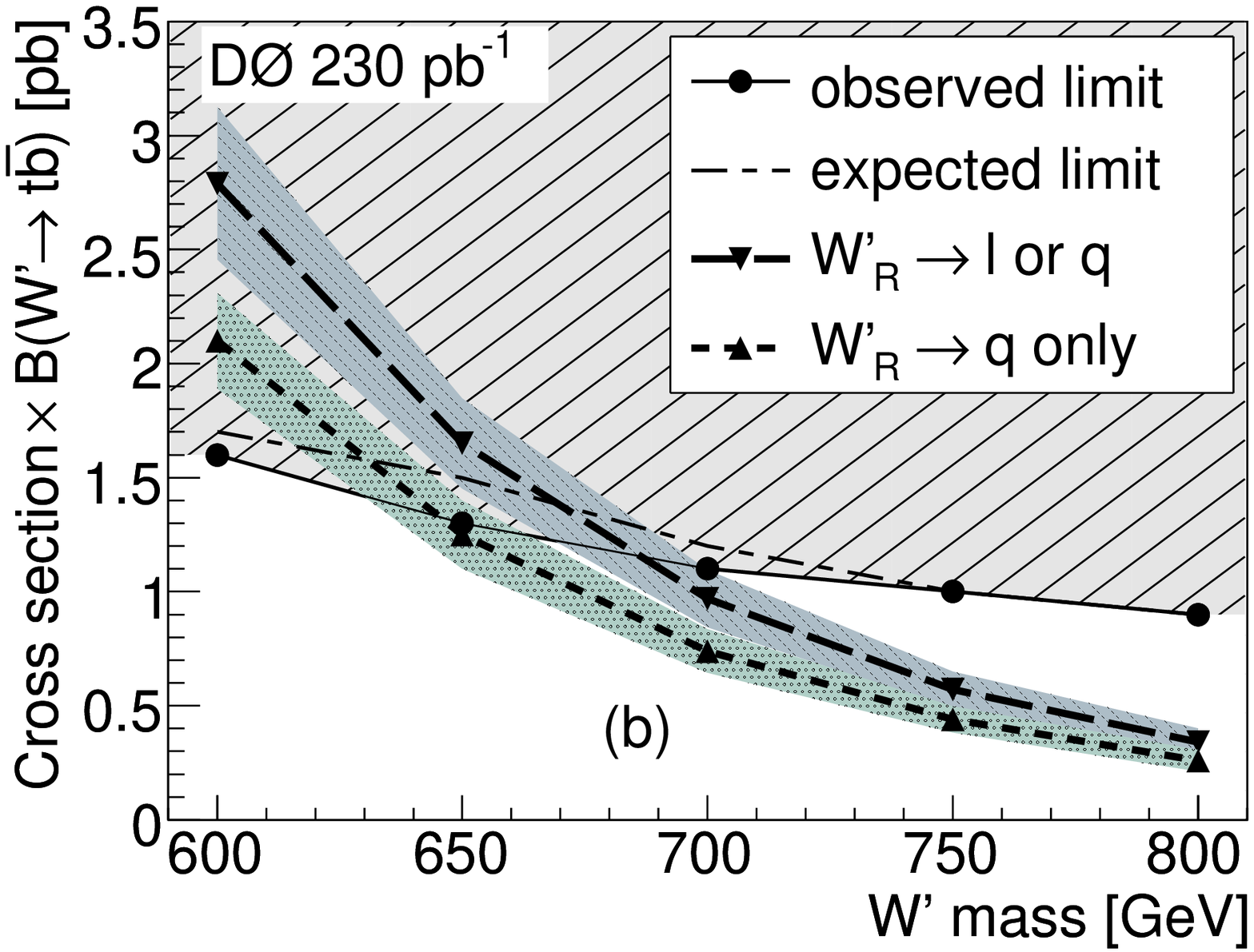}
\end{center}
\caption{Cross section limits at the 95\% confidence level versus the mass of the
$W'$~boson with (a) left-handed couplings and (b) right-handed
couplings.
Also shown are the NLO cross sections according to Table~\ref{tab:xstable}
and the expected limits.
The shaded regions above the circles are excluded by this measurement.}
\label{fig:masslimit}
\end{figure}

% Paragraph 27
%%%%%%%%%%%%%%%%%%%%%%%%%%%%%%%%%%%%%%%%%%%%%%%%%%%%%%%%%%%%%%%%%%%%%%%%%%
% Limits
%%%%%%%%%%%%%%%%%%%%%%%%%%%%%%%%%%%%%%%%%%%%%%%%%%%%%%%%%%%%%%%%%%%%%%%%%%
We combine the electron and muon, single-tagged and double-tagged analysis channels. 
Figure~\ref{fig:masslimit} shows the cross section limits together 
with the cross sections from Table~\ref{tab:xstable} and their uncertainties.

% Paragraph 28
At the 95\% confidence level, the shaded areas above the solid lines are 
excluded by this analysis. The
intersection of the solid line with the lower edge of the uncertainty band 
on the predicted cross section defines the 95\% confidence level
lower mass limit for each model.
Together with the limit from the SM $s$-channel single top quark
search~\cite{Abazov:2005zz}, we thus exclude the presence of a $W'$~boson 
with SM-like left-handed coupling
if it has a mass between 200~GeV and 610~GeV. We also exclude the presence of a
$W'$~boson with right-handed couplings that is allowed to decay to
leptons and quarks (only quarks) if it has a mass between 200~GeV and
630~GeV (670~GeV).
This is the first direct search limit for $W'$~boson production that takes
interference with the SM into account properly. It is also the most
stringent limit in the top quark decay channel of the $W'$~boson.

\section*{Acknowledgments}
We are grateful to Tim Tait for discussions related to this search.
% acknowledgement_paragraph_r2.tex                6/19/06
%
We thank the staffs at Fermilab and collaborating institutions, 
and acknowledge support from the 
DOE and NSF (USA);
CEA and CNRS/IN2P3 (France);
FASI, Rosatom and RFBR (Russia);
CAPES, CNPq, FAPERJ, FAPESP and FUNDUNESP (Brazil);
DAE and DST (India);
Colciencias (Colombia);
CONACyT (Mexico);
KRF and KOSEF (Korea);
CONICET and UBACyT (Argentina);
FOM (The Netherlands);
PPARC (United Kingdom);
MSMT (Czech Republic);
CRC Program, CFI, NSERC and WestGrid Project (Canada);
BMBF and DFG (Germany);
SFI (Ireland);
The Swedish Research Council (Sweden);
Research Corporation;
Alexander von Humboldt Foundation;
and the Marie Curie Program.
%

%---------------------------------------------------------------------
%---------------------------------------------------------------------
%\clearpage


\begin{thebibliography}{99}
\bibitem{fiducial_endnote}
Pseudorapidity is defined as $\eta = - \ln(\tan \frac{\theta}{2})$, 
where $\theta$ is the polar angle with respect to the beam axis,
with the origin at the primary vertex. 
Detector fiducial regions are defined
by detector pseudorapidity  $\eta_{\rm det}$ which is calculated with
the origin at the nominal center of the detector ($z=0$).
\end{thebibliography}
\end{document}